# A Synthetic Texas Power System with Time-Series Weather-Dependent Spatiotemporal Profiles

Jin Lu, *Student Member, IEEE*, and Xingpeng Li, *Senior Member, IEEE*, Hongyi Li, Taher Chegini, Carlos Gamarra, Y. C. Ethan Yang, Margaret Cook, and Gavin Dillingham

***Abstract*—We developed a synthetic Texas 123-bus backbone transmission system (TX-123BT) with spatio-temporally correlated grid profiles of solar power, wind power, dynamic line ratings and loads at one-hour resolution for five continuous years, which demonstrates unique advantages compared to conventional test cases that offer single static system profile snapshots. Three weather-dependent models are used to create the hourly wind power productions, solar power productions, and dynamic line ratings respectively. The actual historical weather information is also provided along with this dataset, which is suitable for machine learning models. Security-constrained unit commitment is conducted on TX-123BT daily grid profiles and numerical results are compared with the actual Texas system for validation. The created hourly DLR profiles can cut operating cost from $8.09 M to $7.95 M (-1.7 %), raises renewable dispatch by 1.3 %, and lowers average LMPs from $18.66 to $17.98 /MWh (-3.6 %). Two hydrogen options—a 200 MW dual hub and a 500 MW hydrogen-energy transmission and conversion system—reduce high-load Q3 daily costs by 13.9 % and 14.1 %, respectively. Sensitivity tests show that suppressing the high-resolution weather-driven profiles can push system cost up by as much as 15 %, demonstrating the economic weight of temporal detail.**

## I. Introduction

Simulation-based studies in power systems frequently utilize publicly available synthetic test cases as benchmark systems due to the scarcity of real system data. These encompass a broad spectrum of research areas, including power system operation and planning [1]-[3], stability and reliability [4]-[5]. Generally, a test case includes all the relevant information for generation, transmission and load. The commonly used test cases are IEEE and CIGRE benchmarks such as the IEEE 73-bus system and CIGRE medium voltage system [6]. Besides these small-scale test power systems, very few large-scale real power system cases are publicly accessible due to the confidentiality of the power industry. To satisfy the need for large-scale cases when real data are unavailable, researchers have developed synthetic grids that mirror actual systems in their electrical characteristics. The Polish 2746-bus system is created based on the real power system of Poland [7]. The synthetic grids utilizing the footprint of the western, northeastern, and eastern U.S. regions have been created, and each grid contains more than ten thousand buses [8]. Most existing test cases provide the technical details for steady-state analysis, such as power flow and/or transient-state analysis such as stability simulation. However, these test cases only provide the data for a certain snapshot in time. The long-term time series system profiles are not provided in these test power system cases. There is a lack of test systems with profiles spanning long continuous periods; such test cases can be used to demonstrate the versatility of renewable production, load and other grid components and enable more comprehensive grid operations and planning studies [9]-[11]. Moreover, the recent trend of applying machine learning (ML) approaches in power systems also requires a test power system with time-series spatio-temporally correlated data over multiple years for ML model training and validation [12]. To bridge the gap, the primary objective of this work is to develop and openly release power system profiles that are bundled with multi-years of hourly wind, solar, dynamic line rating and load profiles that mimic actual large scale power systems. This test case can give the research community a realistic yet computationally manageable platform for operations, planning and machine learning studies that demand long, weather driven time series in addition to large scale network topology. To bridge this gap, we develop a large-scale synthetic test case, the Texas 123-bus backbone transmission (TX-123BT), which is designed to mirror actual systems like Electric Reliability Council of Texas (ERCOT) with detailed weather-dependent grid profiles. It offers a dataset with 5-year time span from 2017 to 2021 at one-hour resolution, enabling enriched analysis and validation in diverse power system studies including ML-related studies. The Texas ERCOT system, with its wide geographic footprint, high renewable penetration, and publicly available historical data, was chosen as the foundation for TX-123BT because it allows the synthetic case to realistically reflect modern grid challenges and conditions.

Power systems will increasingly confront climate-related challenges in the 21st century [13]. However, despite these emerging challenges, the impact of climate and weather on power system performance and potential mitigation strategies has yet to be explored comprehensively. As power systems are transitioning from fossil fuel-based generations to weather-dependent renewable energy generations, current public test cases widely used in academia are no longer effective to support studies targeting at today's power system industry, especially considering a large volume of research papers adopt the emerging machine learning technologies that require large amounts of multi-year grid data. To contribute to this field of

Jin Lu and Xingpeng Li are with the Department of Electrical and Computer Engineering, University of Houston, Houston, TX, 77204, USA (e-mail: jlu28@uh.edu; xingpeng.li@asu.edu).

Hongyi Li and Taher Chegini are with the Department of Civil and Environmental Engineering, University of Houston, Houston, TX, 77204, USA.

Carlos Gamarra, Margaret Cook, and Gavin Dillingham are with Houston Advanced Research Center, Houston, TX, USA.

Y. C. Ethan Yang is with the Department of Civil & Environmental Engineering, Lehigh University, Bethlehem, PA USA.

This work was supported by the Alfred P. Sloan Foundation.

research and facilitate industry advancements, we developed the TX-123BT system with detailed weather-dependent profiles, including wind production, solar production, and dynamic line rating. The spatio-temporal profiles are generated through various weather-dependent models based on the historical weather data from Phase 2 of North American Land Data Assimilation System (NLDAS-2). The scale of the TX-123BT weather-dependent profiles is suitable for environmental condition informed power system studies or data-enriched studies such as machine learning approaches. Furthermore, simulations such as security-constrained unit commitment (SCUC) for large-scale power systems require extensive computational resources [14]. To effectively support research on climate and weather impacts, as well as analyses that cover vast areas and extended timelines, the TX-123BT is specifically designed to encompass only the backbone of the high-voltage transmission network. This focus helps manage computational demands efficiently. Consequently, studies that rely on single-snapshot test cases may underestimate congestion, mis-price energy, and overstate reliability, because they miss the hour-to-hour swings driven by weather. Open-access, multi-year profiles are also vital for the new wave of machine-learning applications that forecast net load, detect anomalies, and optimise market bids. In addition, only time-resolved data allow researchers to evaluate emerging flexibility options—such as dynamic line rating, large-scale battery storage, or power-to-hydrogen schemes—under realistic operating stress. Without this information, the sector cannot fully explore or compare adaptation strategies for a warmer, more variable climate.

The created TX-123BT system, including its network and generator configurations, spatio-temporally correlated grid profiles and the associated historical weather data, has been released with free access [15]-[16]. The main contributions of this paper are as follows,

- A synthetic test case resembling a real-world large geographic area power system (ERCOT) was created and publicly accessible, which has weather-dependent spatio-temporal renewable production and line rating profiles for 5 continuous years at one-hour resolution.
- The test case includes geographic locations of generation and transmission infrastructures. The historical weather and spatio-temporal profiles at these locations are organized as Geographic Information System (GIS) files.
- Both the hourly and daily dynamic line rating (DLR) profiles are provided. The performances of these two DLR techniques are examined and compared.
- The benchmark of the test case is presented by comparing electricity market similarity with actual ERCOT system using the numerical results from SCUC simulations on test case daily profiles.
- By utilizing the weather-dependent renewable production profiles of the test case, the flexibility of hydrogen facilities for high-renewable penetration grid applications is further investigated. The suitable future hydrogen facilities investments are included as additional test case profiles to facilitate other hydrogen and planning related research and studies.

The rest of this work is structured as follows. Section II explains the procedures and workflow to create the TX-123BT test case including generation, transmission and load profiles. Section III presents the generator specifications while section IV presents the weather-dependent renewable production models used for creating renewable production profiles of the TX-123BT. The proposed method for creating the nodal load profiles is described in Section V. Section VI presents the weather-dependent transmission line rating model; the daily and hourly DLR profiles are also summarized. The SCUC simulation results are analyzed in Section VII. The hydrogen integration studies are presented in Section VIII. The conclusions are drawn in Section IX.

## II. System Design Workflow & Profile Scenarios

The aim of this study is to create a power system test case providing both grid configurations and long-term consecutive profiles, instead of a single time snapshot like other test cases. After grid configurations including the network topology and generator parameters are determined, the long-term profiles are created based on the configurations and parameters of the transmission and generation components. The workflow of TX-123BT creation is illustrated in Fig. 1.

Firstly, we determine the infrastructure details for power system facilities, including the locations and parameters for transmission lines, substations, thermal generators and renewable power plants. The geographic distribution of the TX-123BT transmission network is based on the 345kV high voltage transmission lines in Texas [8] and is shown in Fig. 2. For thermal power plants, the output capacity, ramping rate, and generation costs are crucial for dispatching decisions to ensure reliability while minimizing costs. For renewable power plants such as wind farms, the configurations including locations, wind turbine blade length, and the overall plant capacity are required for creating the wind production profiles for a long period. We consider their geographic distribution and reasonable capacity range of generators with different fuel types based on the ERCOT fuel mix and EIA generation data. In TX-123BT, the generator and transmission network are located coordinately similar to the real actual power systems.

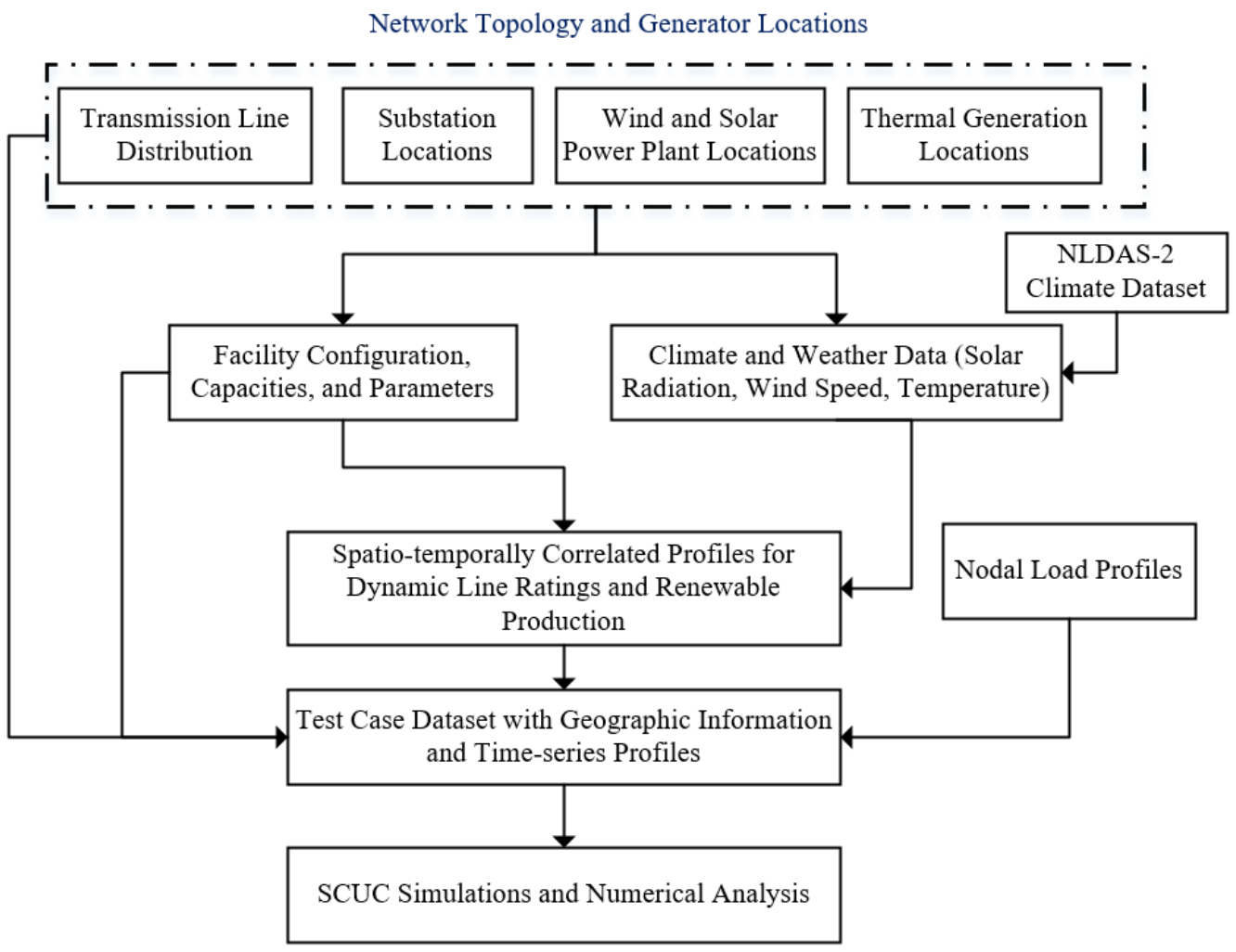

Fig. 1. The workflow of TX-123BT test case creation.

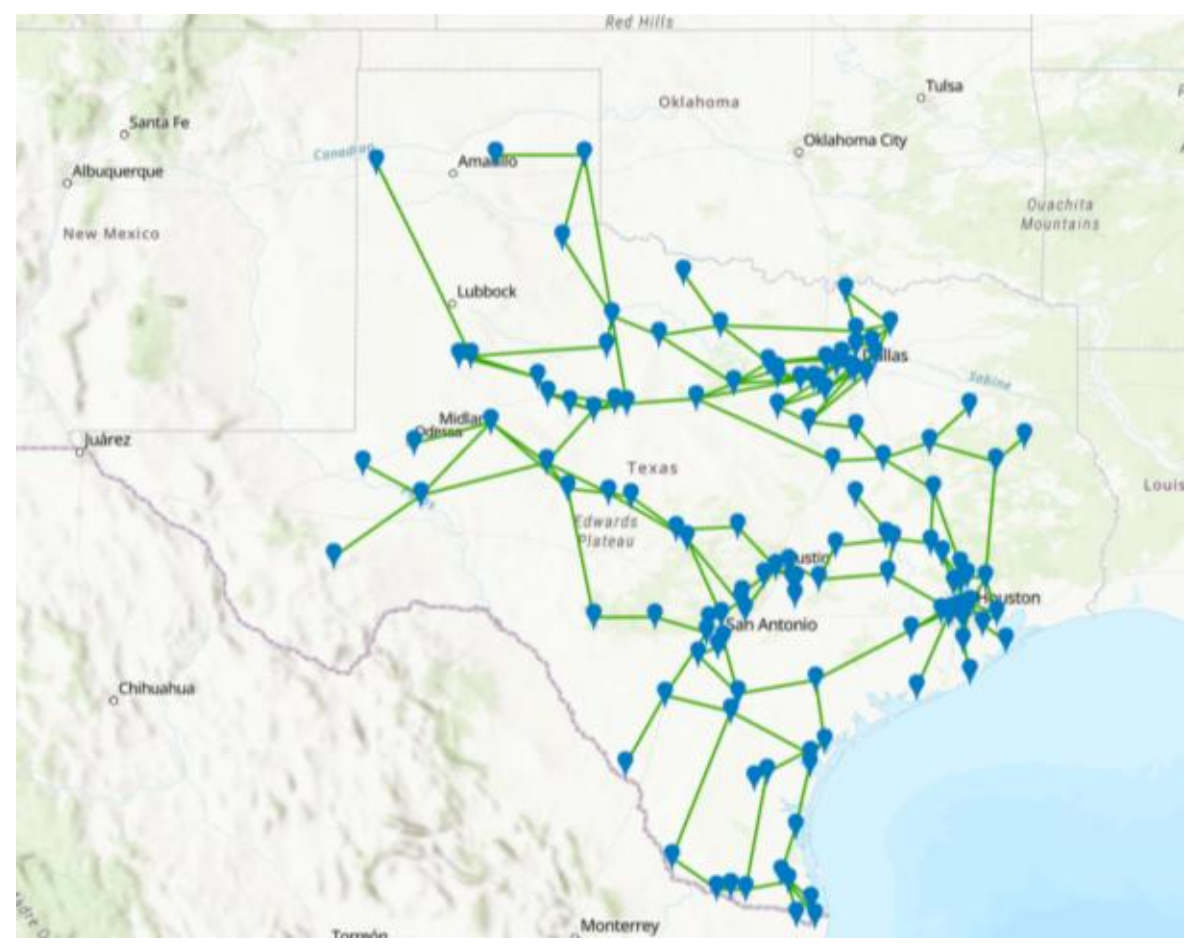

Fig. 2. Illustration of the 123-bus transmission network topology.

Secondly, we develop weather-dependent profiles based on facility configurations and the historical weather data at the locations of these facilities. Both hourly and daily line ratings are calculated based on the transmission line conductor types, sizes, reactance, as well as the temperature and wind speed data on two terminals of the lines. Wind production is calculated based on the wind turbine parameters such as blade length, cut-in and cut-out wind speed and the estimated wind speed at the wind turbine height. Solar production is calculated based on the solar panel parameters and the estimated solar radiation for the effective wavelength on the panel.

Thirdly, we create the time-sequential load profiles based on the ERCOT historical load data on different weather zones. Then, we organized the above-mentioned profiles and made it ready for power system both short-term and long-term analysis such as SCUC and long-term transmission planning.

To validate the designed TX-123-BT system, we ran SCUC for the daily profiles in the 5 years and compared the obtained electricity prices with the actual ERCOT electricity market. We also evaluated the transmission network congestion of the TX-123BT during peak load scenarios and analyzed the grid performance with hourly and daily line ratings.

The TX-123BT with high resolution renewable production profiles is utilized to study the configurations and performance of hydrogen facilities in high renewable penetration grids. It is also an example to demonstrate the necessity of the TX-123BT spatio-temporally correlated profiles. Specifically, we have created the TX-123BT profiles coupled with hydrogen hubs and Hydrogen Energy Transmission and Conversion System (HETCS) [48] separately. The grid operation conditions with different hydrogen solutions are evaluated and compared, with the help of detailed renewable production profiles provided by TX-123BT.

After completing the workflow and profile generation, TX-123BT emerges as a 123-bus, 345 kV backbone representation of ERCOT. It embeds 292 generating units, totaling 103.4 GW of name-plate capacity: 113 natural-gas units (56.0 GW), 82 wind farms (24.5 GW), 72 utility-scale solar plants (2.48 GW), 13 coal plants (14.8 GW), 10 hydro stations (0.50 GW) and two nuclear units (5.14 GW) . Each generator is placed in the eight ERCOT weather zones so that the system - wide and zonal fuel mixes closely match 2019 ERCOT statistics. Five full years of hourly, bus-level profiles (2017–2021; 43 800 h per series) are supplied for wind, solar, dynamic line ratings and load, giving a high-resolution spatio-temporal data set for both conventional studies and machine-learning applications. The transmission network preserves the major north–south and west–east 345 kV corridors (Fig. 2) and is sized to keep SCUC runs tractable while still reproducing realistic congestion patterns and locational price spreads.

The TX-123BT system, as currently provided, is intended for steady-state and economic analyses (e.g. unit commitment, production cost, planning studies). While TX-123BT provides extensive time-series profiles and a realistic network topology, it is important to note its limitations. i) The test case focuses on the transmission-level steady-state operations and excludes dynamic system parameters (e.g., generator inertia, transient stability models); thus, analyses of frequency stability or inertial response are beyond its current scope without additional data; ii) The model is confined to the 345 kV backbone network, so it does not capture distribution networks or lower voltage details, which may be relevant for certain studies; iii) All profiles are hourly resolution; sub-hourly dynamics or fast transient events would require higher-resolution data. These limitations should be considered when employing TX-123BT for research.

## III. Conventional Generation Profiles

### A. Generation Fuel Mix

Major generation fuel types in the ERCOT system are natural gas, wind, coal, nuclear, and solar per [20]-[21]. The generation fuel composition may vary in different regions of ERCOT. For example, most wind generators are in the northwest of Texas due to the wind source distribution. The TX-123BT is created to have a very similar system-wide as well as region-wide generation fuel mix with the actual ERCOT system. Based on the generation characteristics of various fuel types for different weather zones and the whole ERCOT [22]-[24], fuel types of the generators in the TX-123BT system are assigned accordingly and the generator's power capacity is within the capacity range of the corresponding type of generators. Based on the data provided by Energy Information Administration (EIA) [22], ten hydro power plants, each of which is over 10 MW, are also added to the generation profiles.

The statistical data of TX-123BT generation profiles are shown in Tables I-II. The system-wide generation fuel mix (last row in Table II) is similar to the actual fuel mix provided by ERCOT [25]. It is worth noting that the renewable generation capacity has grown rapidly in ERCOT. We have adjusted the wind and solar capacity to match the actual ERCOT renewable capacity in 2019.

### B. Conventional Generator Costs & Operation Parameters

#### 1) Coal & Natural Gas Generator

The quadratic function (1) is used to model the thermal power plant's operation cost ($c_g$). The coefficients $C_0$, $C_1$ and $C_2$ of typical coal generators and natural gas generators are determined per [26]. The generator startup cost ($c_g^{SU}$) can be calculated using (2).

$$c_g = C_0 + C_1 * P + C_2 * P^2 \tag{1}$$

$$c_g^{SU} = \eta_g^{SU} * P_g^{Max} * C^F + P_g^{SU} * C_g^O \tag{2}$$

where $P_g^{Max}$ is the generator active power capacity, $\eta_g^{SU}$ is the startup fuel per unit capacity, $C^F$ is the fuel price, $P_g^{SU}$ is the startup power and $C_g^O$ is another startup cost related to the required startup power.

The coal price used in creating the TX-123BT system is 1.78 \$/MMBtu based on EIA [26]. The annual average natural gas price in Texas is 2.29 $\$/Kft^3$. As the natural gas heat content is set to be 1000 $Btu/ft^3$ [27], then the natural gas price becomes 2.29 $\$/MMBtu$. Based on the above information, the total startup cost is calculated for the coal and natural gas generators in the TX-123BT test system. Tables III show the cost range for coal generators in TX-123BT.

In addition, the shutdown costs are calculated following [23]. The ramping rate, minimum off time, and maximum on time of coal and natural gas generators are obtained per [24]. The startup time and shutdown time are obtained from [29].

TABLE I Total Number of Different Fuel Type Generators

| Weather Zone | Natural Gas | Wind | Coal | Solar | Nuclear | Hydro |
|---|---|---|---|---|---|---|
| Coast | 31 | 0 | 2 | 8 | 1 | 0 |
| East | 15 | 0 | 2 | 0 | 0 | 0 |
| Far West | 4 | 11 | 0 | 17 | 0 | 0 |
| North | 4 | 9 | 1 | 6 | 0 | 1 |
| North Central | 14 | 10 | 4 | 9 | 1 | 1 |
| South | 11 | 18 | 1 | 3 | 0 | 1 |
| South Central | 29 | 0 | 3 | 23 | 0 | 6 |
| West | 5 | 34 | 0 | 6 | 0 | 1 |
| Total | 113 | 82 | 13 | 72 | 2 | 10 |

TABLE II Total Capacities (MW) of Different Fuel Type Generators

| Weather Zone | Natural Gas | Wind | Coal | Solar | Nuclear | Hydro |
|---|---|---|---|---|---|---|
| Coast | 11925 | 0 | 2373 | 37 | 2709 | 0 |
| East | 6385 | 0 | 3089 | 0 | 0 | 0 |
| Far West | 3892 | 2341 | 0 | 1760 | 0 | 0 |
| North | 2236 | 3990 | 802 | 256 | 0 | 80 |
| North Central | 12093 | 1277 | 4529 | 42 | 2431 | 42 |
| South | 7480 | 6717 | 410 | 97 | 0 | 31 |
| South Central | 11075 | 0 | 3560 | 159 | 0 | 287 |
| West | 920 | 10157 | 0 | 131 | 0 | 58 |
| Total | 56006 | 24480 | 14761 | 2480 | 5139 | 498 |

TABLE III Costs for Various Scale Coal Power Plants

| Capacity (MW) | C0 (\$/h) | C1 (\$/h/ MW) | C2 (\$/h/ MW$^2$) | Startup (\$/Installed MW) | Shutdown (\$/Installed MW) |
|---|---|---|---|---|---|
| 0-75 | 0-238 | 18.28-19.98 | 0.0016 | 80-380 | 8-38 |
| 75-150 | 238-745 | | | | |
| 150-350 | 745-1213 | | | | |
| >350 | 1213-3043 | | | | |

*2) Nuclear Power Plants*

Texas has two nuclear power plants. Per expense of nuclear power plants [30], we assume the operation cost of the nuclear power plants is 17.44 \$/MWh. The nuclear power plants are generally online most of the time. Most nuclear power plants require more than 12 hours to reach full operation, and they can ramp up or down in the load following mode [31].

*3) Hydroelectric Power Stations*

The average operation cost of hydroelectric power plants is 12.3 \$/MWh per [32]. Hydroelectric power stations can ramp up rapidly and require much less time to startup and shutdown. As an example, hydroelectric power plants provided flexible ramping during the generation shortage in Texas in February 2021 [33]. Thus, we assume the startup/shutdown time, minimum on/off time of hydroelectric power stations are zeros.

## IV. Time-Series Weather-dependent Spatio-temporally Correlated Renewable Power Production Profiles

### *A. Weather-dependent Wind Model and Production Profiles*

The TX-123BT renewable production profiles depend on the weather data in at the plant locations, and we obtain these data from NLDAS-2 [17]-[18]. The NLDAS-2 climate dataset includes the shortwave/longwave solar radiation, air temperature and wind speed near the ground surface at one-hour resolution. Since this paper aims to create the time-series profiles that resemble the ERCOT in the 5-year period from 2017-2021, the historical weather data for all the hours in the period are extracted.

In the ERCOT system, the full capacity up to the high sustained limit (HSL) of wind generation resources is considered available to be dispatched in the reliability unit commitment (RUC) [34]. However, ERCOT only provides limited data for the HSL in the current operating plan (COP), which is not the actual HSL. Instead of determining the wind hourly HSL for daily grid operations, we use the wind hourly production data [35] to create the wind profiles.

The available wind power output can be calculated using (3)-(4) when wind speed is between the cut-in speed and rated-speed. $V$ is the wind speed, $\rho$ is the air density, $C_p$ is the wind turbine efficiency, $A$ is the turbine blade swept area, and $D$ is the turbine diameter. For all the wind turbines in the TX-123BT system, the cut-in speed is set to 3.5m/s, the rated speed is set to 13m/s, and the cut-out speed is set to 25m/s.

$$P^{W,Max} = \frac{1}{2} \cdot \rho \cdot A \cdot V^3 \cdot C_p \tag{3}$$

$$A = \frac{\pi}{4} \cdot D^2 \tag{4}$$

The extracted historical wind speed data in NLDAS-2 is the measured wind speed at 10m height and most wind turbines on land are about 80m high [36]. To determine wind speed at 80m, we use log wind profile [37] to estimate the wind speed at the wind turbines' height. The equation for the log wind profile is shown below.

$$u(z_2) = u(z_1) \cdot \frac{ln((z_2 - d)/z_0)}{ln((z_1 - d)/z_0)} \tag{5}$$

The wind speeds at two different heights $z_1$ and $z_2$ are represented by $u(z_2)$ and $u(z_2)$ respectively. Zero-plane displacement $d$ is the height that zero-wind speed is achieved because of flow obstacles. It can be approximated as 2/3 to 3/4 of the average height of the obstacles [37]. Roughness length $z_0$ is the roughness of the surface. Based on the terrain of Texas, $z_0$ is set to 0.3 and $d$ is set to 6. The wind speed at 80m is about 2.13 times of wind speed at 10m.

To create more practical weather-dependent wind production profiles, the capacities and geographic locations of the wind power plants in TX-123BT should be close to the actual ERCOT system. Thus, the wind farm capacities in TX-123BT are adjusted to match the actual ERCOT wind generation using a least square method, as described by (6)-(11).

$$min \sum_{h}^{H} (\sum_{i}^{N^W} p_{i,h}^{W,Case} - p_h^{ERCOT})^2 \tag{6}$$

$$P_{i,h}^{W,Case} = k_{i,h}^W \cdot C_i^W \cdot V_{i,h}^3 \quad \forall\, i \in N^W,\, h \in H \tag{7}$$

$$k_{i,h}^W = k_{i,h+24}^W \quad \forall\, i \in N^W ,h \in H \tag{8}$$

$$-E_k^{max} \le k_{i,h}^W - k_{i,h-1}^W \le E_k^{max} \;\forall\, i \in N^W, h \in H \tag{9}$$

$$-E_c^{max} \le C_i^W - C_i^{W0} \le E_c^{max} \quad \forall\, i \in N^W, h \in H \tag{10}$$

$$C_i^W \ge 0 \quad \forall\, i \in N^W \tag{11}$$

The least square method can adjust the wind power plant capacities in the TX-123BT to minimize the square error between the wind production of the TX-123BT and the ERCOT per (6). $p_{i,h}^{W,Case}$ is the power output of wind farm *i* in hour *h* in the TX-123BT. $p_h^{ERCOT}$ is the total wind output power of the actual ERCOT system in hour *h*. Hour *h* is an hour in the period. The aggregated wind power production in a wind farm is related to the adjusted capacity of wind farm $C_i^W$ and the wind speed $V_{i,h}$ per (7).

The wind turbine coefficient *k* is a comprehensive coefficient considering various factors including the wind direction and wind turbine efficiency. The wind turbine coefficient *k* is assumed to be a constant for a specific wind turbine for each hour of the day per (8). The changing magnitude of *k* is limited to $E_k^{max}$ over two consecutive hours. Besides, the adjustment of the capacity for each wind farm in TX-123BT is less than the maximum error $E_c^{max}$ per (10). The adjusted wind capacity should be non-negative per (11). The least square method can find the most realistic wind turbine coefficients and capacities for wind farms in the TX-123BT system.

Fig. 3 compares the synthetic hourly wind-production profiles with the actual ERCOT data for 2019. The mean hourly wind power profile (within a day), averaged over 365 days in 2019, is compared to the actual ERCOT hourly statistics in Fig. 4. The hourly wind power production from seven wind farms at bus 119 on January 3, 2019, is illustrated in Fig. 5. According to the comparison in Fig. 4, we can conclude that the created wind production profiles are very similar to the actual situation during the daytime. However, there is a bias remained during night hours. This bias originates from the nocturnal boundary layer regime common in Texas, where surface cooling decouples the rotor from stronger winds aloft and an elevated low level jet (LLJ) forms at 200–400 m. A fixed shear exponent therefore under predicts or over predicts hub height wind speed depending on the LLJ core height.

### B. Weather-dependent Solar Model and Production Profiles

A five-parameter single diode equivalent circuit is commonly used and suitable for PV cell, module and array [38]. In [39], the operation condition variables (temperature, radiation, and air mass) are used in a five-parameter equation. Since we are mainly interested in the maximum available solar power output at different radiation and temperature, we can calculate the maximum power point using (12) [40]-[42].

$$P_{mp} = \frac{E_e}{E_0} \cdot P_{mp0} \cdot [1 + \gamma \cdot (T_c - T_0)] \tag{12}$$

where $P_{mp}$ is the maximum power output for the certain operation condition. $E_e$ and $T_c$ are the effective radiation and temperature on the solar cells respectively. $P_{mp0}$ is the maximum power output at the standard testing condition (STC). $E_0$ and $T_0$ are the radiation and temperature at STC respectively. $\gamma$ is the temperature coefficient that indicates the influence of the temperature on solar power transfer efficiency.

The NLDAS-2 provides the historical data for shortwave and longwave solar radiation flux downwards. Based on the widely used solar panel's spectral response range, we use the radiation flux downwards to estimate the effective solar radiation on the solar panels. We also estimate the solar cell temperature using the ambient air temperature at the corresponding solar panel.

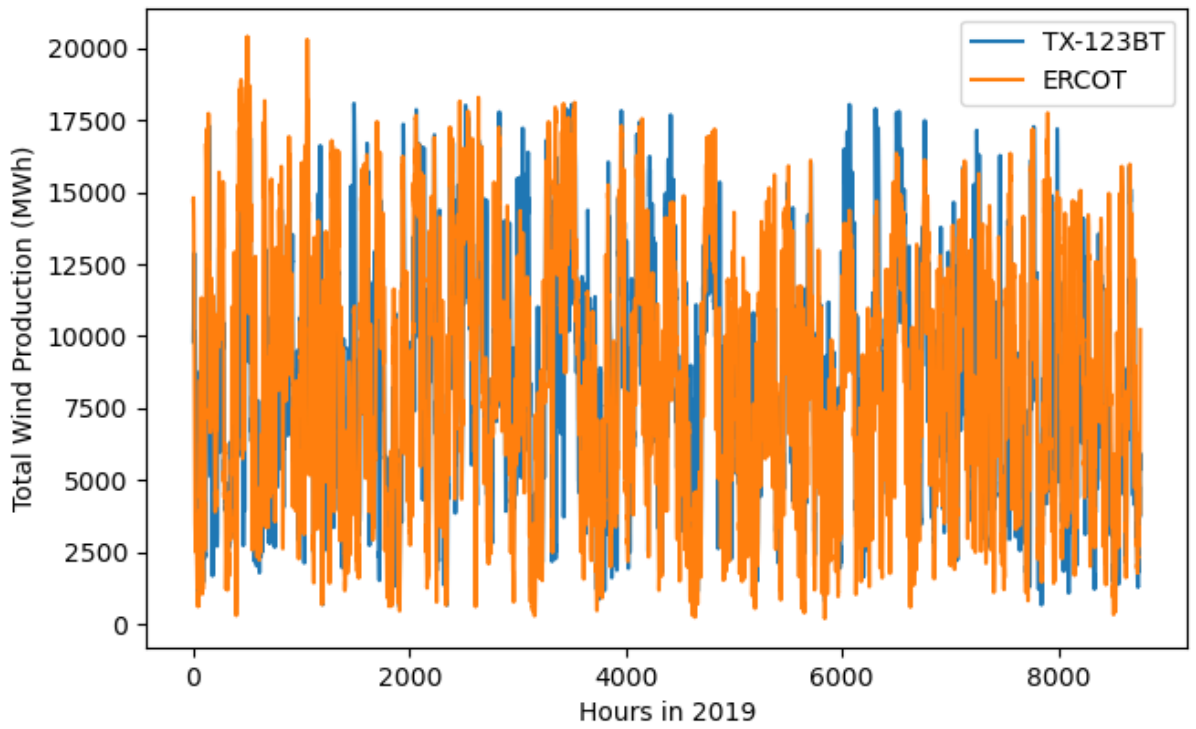

Fig. 3. Wind power production for all hours in 2019.

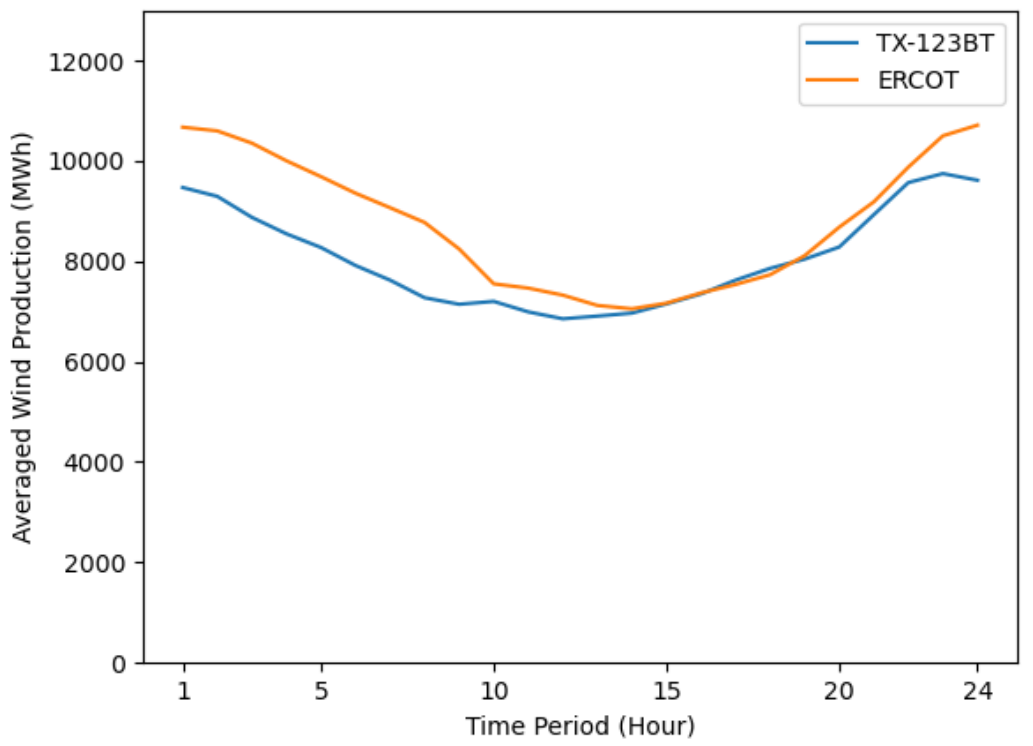

Fig. 4. Hourly wind power profiles comparison.

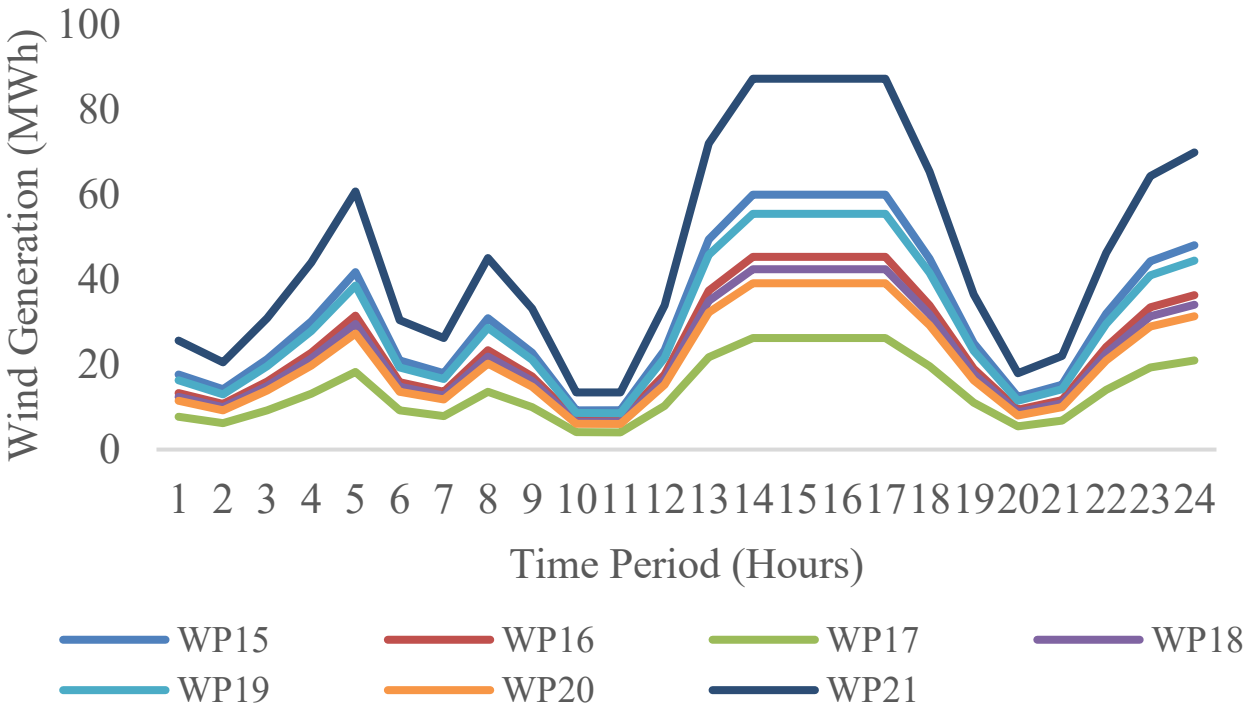

Fig. 5. Output power of seven wind plants (WP) on January 3, 2019.

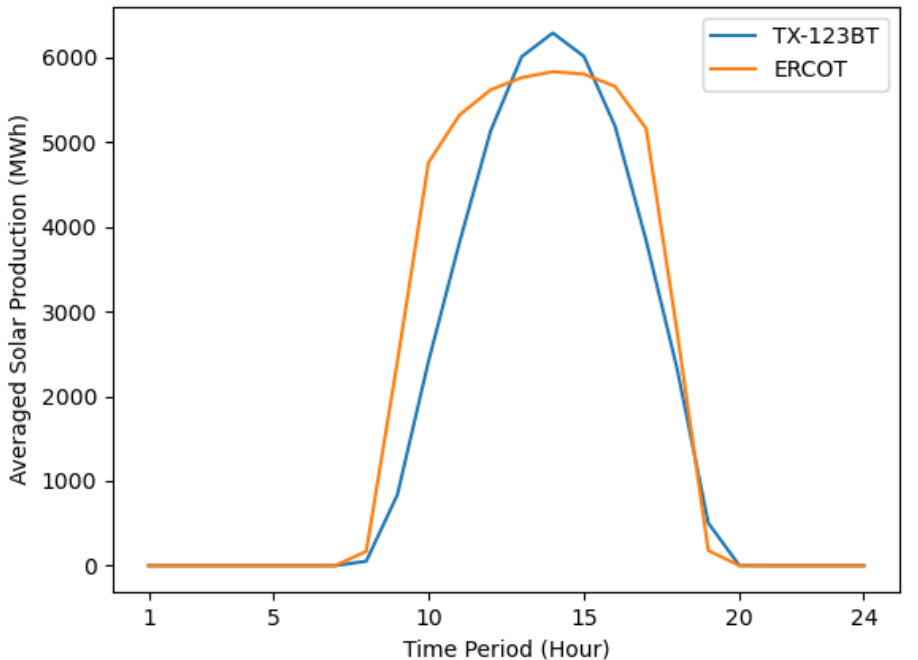

Fig. 6. Averaged hourly solar power production in Quarter 1.

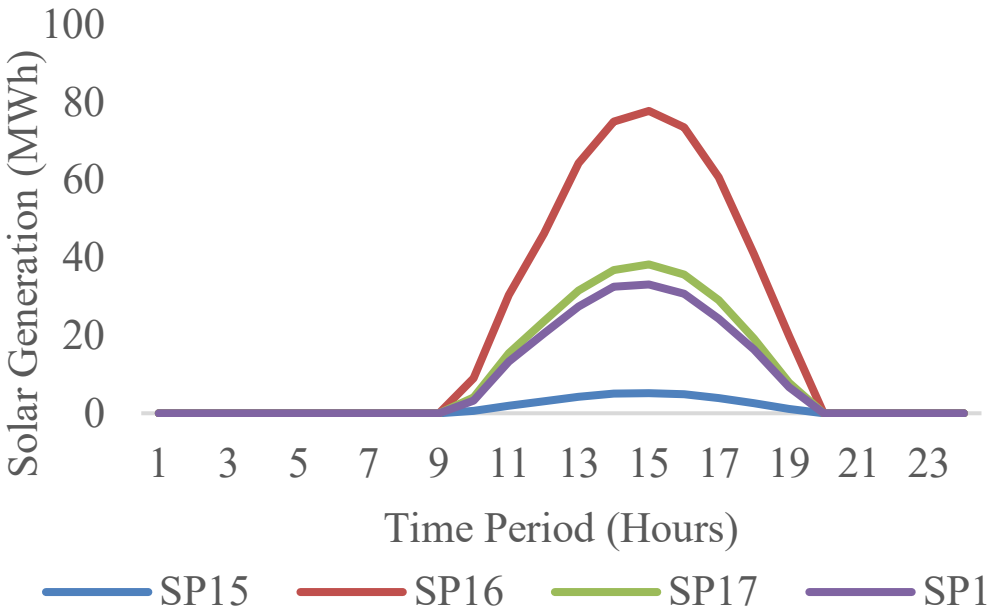

Fig. 7. Hourly power output of four solar plants (SP) on January 1, 2019.

Based on the processed weather data and solar power production model, the solar power production for all the solar farms in the TX-123BT system is calculated. The system-wide hourly solar production of the TX-123BT is compared with ERCOT solar production in 2022 (the historical data is not accessible). The hourly solar productions averaged over all the days in Quarter 1 for the synthetic TX-123BT system and the actual ERCOT system are shown in Fig. 6. The ERCOT curve is noticeably wider (starting earlier in the morning and tapering off later in the evening) than the TX-123BT curve. This is expected, since most utility-scale solar farms in ERCOT employ single-axis tracking [41], which tilts panels from east to west to follow the sun. Such tracking broadens the production window. In our dataset, however, we assumed fixed-tilt PV panels, leading to a narrower generation profile concentrated around noon. The hourly solar power production for four solar farms in TX-123BT is shown in Fig. 7.

## V. Creation of Load Profiles

The ERCOT historical load data [43] includes the hourly total load in each weather zone for all hours in different years. The hourly load profiles at the nodal level are created for the TX-123BT system such that they match the total zonal load amount in each weather zone for each hour from 2017 to 2021. Part of the load profiles are illustrated in Table IV.

TABLE IV Sample Load Profiles for Different Hours on December 31, 2019

| Bus Number | Hour 1 | Hour 5 | Hour 9 | Hour 13 | Hour 17 | Hour 21 |
|---|---|---|---|---|---|---|
| Bus 1 | 96.51 | 102.39 | 118.19 | 91.76 | 84.25 | 96.28 |
| Bus 2 | 97.86 | 103.56 | 116.47 | 95.4 | 87.29 | 97.41 |
| Bus 3 | 171.3 | 172.01 | 173.67 | 161.79 | 158.4 | 165.91 |
| Bus 4 | 198.59 | 202.18 | 231.79 | 213.28 | 208.4 | 212.48 |
| Bus 5 | 25.58 | 24.5 | 28.54 | 28.53 | 28.73 | 28.21 |
| Bus 6 | 606.36 | 641.1 | 754.35 | 608.1 | 555.7 | 613.43 |
| Bus 7 | 42.61 | 45.05 | 53.01 | 42.73 | 39.05 | 43.11 |

## VI. Weather-dependent Transmission Line Rating

The transmission line thermal capacity in the three-phase system can be calculated given the line ampacity and line voltage. The IEEE Std 738-2012 [44] is used to calculate ampacity of lines at different temperature, solar radiation and wind speed conditions. The detailed calculation is described by (13)-(20).

$$q_c + q_r = q_s + I^2 \cdot R(T_{avg}) \tag{13}$$

$$I = \sqrt{\frac{q_c + q_r - q_s}{R(T_{avg})}} \tag{14}$$

$$q_{c1} = K_{angle} \cdot [1.01 + 1.35 \cdot N_{Re}^{0.52}] \cdot k_f \cdot (T_s - T_a) \tag{15}$$

$$q_{c2} = K_{angle} \cdot 0.754 \cdot N_{Re}^{0.6} \cdot k_f \cdot (T_s - T_a) \tag{16}$$

$$K_{angle} = 1.194 - \cos(\varphi) + 0.194 \cdot \cos(2\varphi) + 0.368 \cdot \sin(2\varphi) \tag{17}$$

$$N_{Re} = \frac{D_0 \cdot \rho_f \cdot V_w}{\mu_f} \tag{18}$$

$$q_r = 17.8 \cdot D_0 \cdot \varepsilon \cdot [(\frac{T_s + 273}{100})^4 - (\frac{T_a + 273}{100})^4] \tag{19}$$

$$q_s = \alpha \cdot Q_{se} \cdot \sin(\theta) \cdot A' \tag{20}$$

$$\theta = \cos^{-1}[\cos(H_c) \cdot \cos(Z_c - Z_l)] \tag{21}$$

Equation (13) is the heat balance equation of the conductor. $q_c$ is the convective heat loss. $q_r$ is the radiated heat loss rate. $q_s$ is the rate of solar heat gain. $I$ is the current in the conductor and $R(T_{avg})$ is the conductor resistance at temperature $T_{avg}$, which is the average temperature in the conductor. (13) can be transformed into (14), which can be used to calculate the current in the conductor at the conductor maximum temperature. In (15)-(17), $q_{c1}$ and $q_{c2}$ are the forced convection and the higher value of $q_{c1}$ and $q_{c2}$ will be used as the value of $q_c$. $N_{Re}$ is the dimensionless Reynolds number. $K_f$ is the thermal conductivity of air. $T_s$ is the conductor surface temperature, and $T_a$ is the ambient temperature. $K_{angle}$ is wind direction factor and $\varphi$ is the angle between the wind direction and the conductor axis. In (18), the Reynolds number $N_{Re}$ is directly proportional to the conductor diameter $D_0$ and wind velocity $V_w$. Air density $\rho_f$ and the dynamic viscosity of air $\mu_f$ are calculated at the mean film temperature of the conductor boundary layer. In (19), the radiated heat loss is related to the diameter of the conductor $D_0$, the emissivity $\varepsilon$, the conductor surface temperature $T_s$, and ambient temperature $T_a$. The rate of solar heat gain can be calculated by (20). $\alpha$ is the solar absorptivity. $Q_{se}$ is total solar and sky radiated heat intensity corrected for the elevation. $A'$ is the projected area of the conductor. $\theta$ is the effective angle of incidence of the sun's rays. In (21), $\theta$ is determined by the altitude of the sun $H_c$, the azimuth of the sun $Z_c$, and the azimuth of the line $Z_l$.

There are three types of aluminium conductor steel reinforced (ACSR) conductors used for the transmission lines in the TX-123BT: Kiwi, Bobolink and Finch. Different types of ACSR conductors have different conductor diameters and resistances versus temperature characteristics. The linear approximation (22) determines the conductor resistance at a cer-tain temperature. $R(T_{high})$ and $R(T_{low})$ are the conductor resistance at temperature $T_{high}$ and $T_{low}$ respectively.

$$R(T_{avg}) = [\frac{R(T_{high}) - R(T_{low})}{T_{high} - T_{low}}] \cdot (T_{high} - T_{low}) + R(T_{low}) \tag{22}$$

Although the extracted historical weather data have detailed nodal information at one-hour resolution, they do not perfectly meet the needs of transmission line calculation. Several assumptions are made as follows. First, the line ambient temperature is assumed to be the same as the temperature 2 meters above ground. Second, since most long-distance transmission lines are overhead lines and the transmission towers are generally 55-150 feet (16.8m-45.72m), the wind speed at the transmission line's height is estimated using the aforementioned log wind profile method. Third, the angle between the wind direction and the transmission line is assumed to be 45-degrees. The wind speed perpendicular to the conductor $V_w$ can be calculated using (23)-(24). In (23), $V_z$ and $V_m$ is the zonal and meridional wind speed extracted from NLDAS-2. $V_{wind}$ is the composite speed, and $\psi$ is the angle between the wind turbine blade and wind direction.

$$V_{wind} = \sqrt{{V_z}^2 + {V_m}^2} \tag{23}$$

$$V_w = V_{wind} * sin(\psi) \tag{24}$$

The total heat intensity corrected for elevation $Q_{se}$ is calculated using (25)-(27) per the IEEE Std 738-2012.

$$Q_s = A + B \cdot H_c + C \cdot {H_c}^2 + D \cdot {H_c}^3 + E \cdot {H_c}^4 + F \cdot {H_c}^5 + G \cdot {H_c}^6 \tag{25}$$

$$Q_{se} = K_{solar} \cdot Q_s \tag{26}$$

$$K_{solar} = A + B \cdot H_e + C \cdot {H_e}^2 \tag{27}$$

In (24)-(26), $Q_s$ is the total heat flux density by a surface at sea level. $K_{solar}$ is the elevation corrective factor. *A*, *B*, *C*, *D*, *E*, *F*, and *G* are polynomial coefficients.

Instead of (25)-(27) which is the IEEE Std 738-2012 method of computing clear-sky solar heat flux at sea level and applying an elevation correction, in this work, we calculate the total solar heat flux $Q_s$ using the NLDAS-2 radiation data. Specifically, $Q_s$ is obtained by summing the downward shortwave and longwave radiation components (with no ground reflection), as given by (28).

$$Q_s = Q_{short} + Q_{long} \tag{28}$$

In the calculation, the environmental parameters are determined based on the actual Texas conditions. The altitude and azimuth of the sun at noon are used in the calculation. The altitude of the sun $H_c$ is calculated based on the average latitude of Texas which is 30.5° N. The elevation of the conductor above the sea level $H_e$ is set to Texas average elevation. For ACSR transmission lines, the common continuous operational maximum temperature is 90°C. The parameters for the Texas line ampacity calculation are listed in Table V.

TABLE V Some Input Data for Texas Line Ampacity Calculation

| Notation | Explanation | Value |
|---|---|---|
| $H_c$ | The altitude of the sun | 30.5 |
| $Z_c$ | The azimuth of the sun | 180 deg |
| $H_e$ | The elevation of conductor above sea level | 520 m |
| $\varepsilon$ | The emissivity | 0.8 |
| $\alpha$ | Solar absorption | 0.8 |
| $\mu_f$ | The air viscosity | 2.04e-5 |
| $T_{film}$ | Average temperature of the boundary layer | 70 °C |
| $K_f$ | The thermal conductivity of air | 0.0295W/(m-°C) |
| $T_c$ | The conductor maximum temperature | 90 °C |

Dynamic line rating is an effective strategy in power system operations to fully utilize the available transmission capacity of the lines under various environmental conditions. In this paper, we have created two profiles using the daily DLR and the hourly DLR, respectively. The daily DLR profile has the same fixed line ratings for the entire day, which is used by many power system operators. We use the highest hourly temperature, solar radiation, and the lowest hourly wind speed as the environmental values in the daily line rating calculation. The daily line rating profiles are calculated for all the days in the 5-year period. The daily thermal ratings of line 15 during 2019 are shown as an example in Fig. 8.

The hourly line rating method captures the suitable line ratings for each hour in the daily operation. Each hour's temperature, solar radiation, and wind speed are used in the line rating calculation for the corresponding hour. Hourly line ratings are higher than the daily line rating in most instances. Hence, using hourly line rating can reduce operational costs and improve system operational efficiency. The hourly line ratings are calculated for all the hours in the 5-year period. The hourly ratings of line 15 for different quarters are shown in Fig. 9.

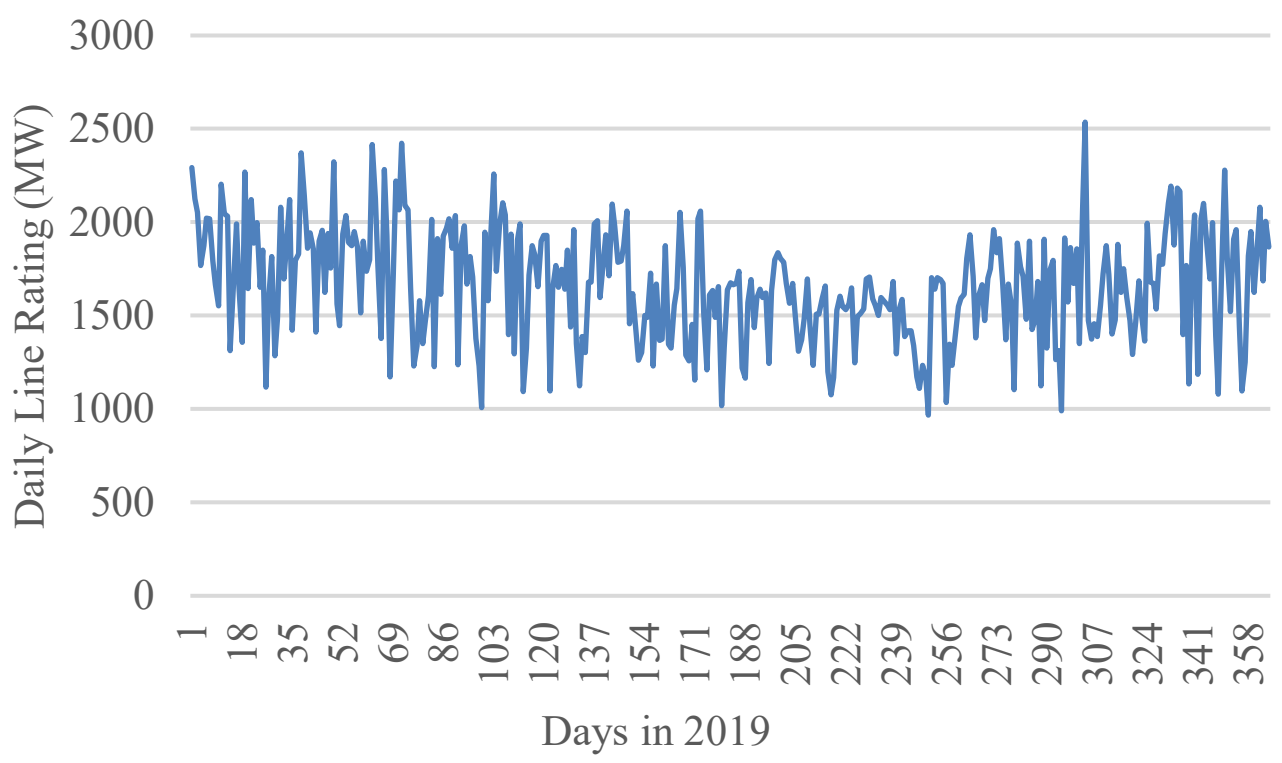

Fig. 8. The daily thermal ratings of line 15 during the year 2019.

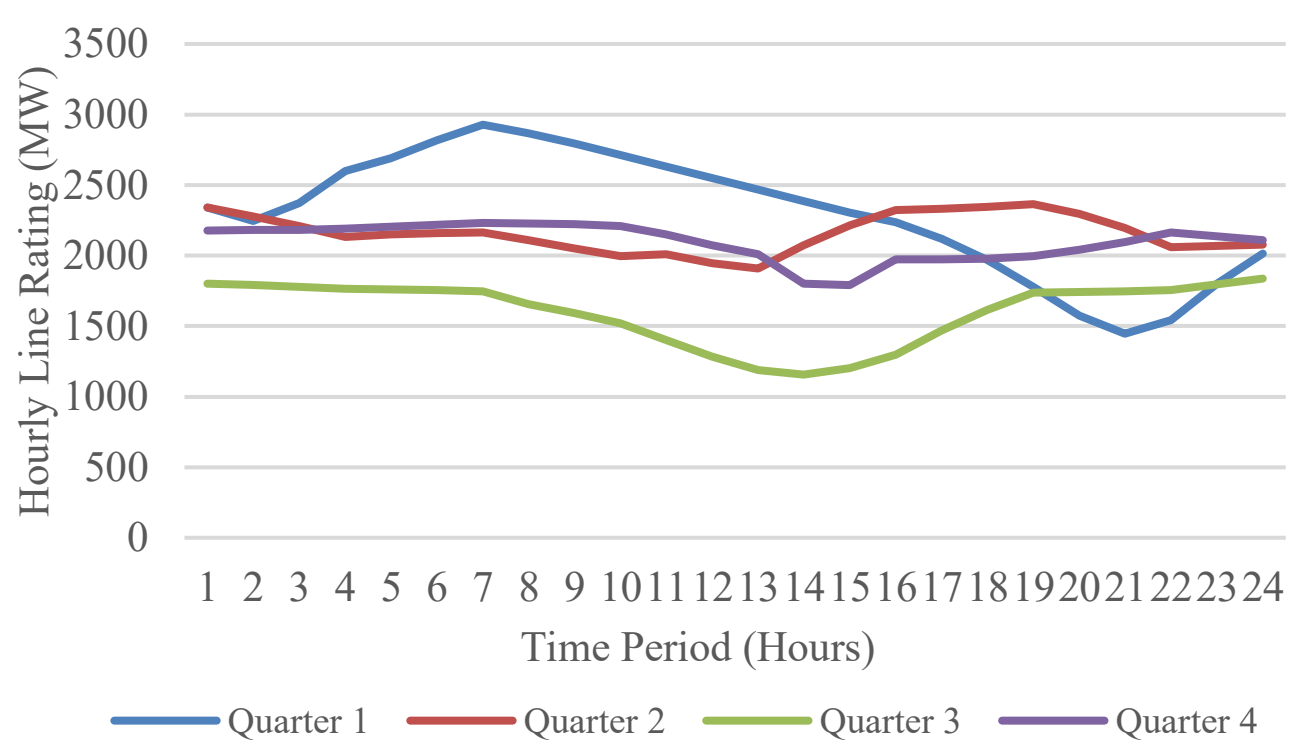

Fig. 9. Hourly ratings of line 15 in four typical days for different quarters.

## VII. SCUC Simulation and Analysis

### *A. SCUC Model*

To analyze the daily operational conditions of the created TX-123BT system, a standard SCUC model in [45] is used and simulation is conducted on the TX-123BT. The objective of the daily operational model is to minimize the total cost of the power system, which includes the generator operation cost, no-load cost and startup cost. The generator related constraints such as maximum output, reserve, and ramping constraints are included. The line flow equation and nodal power balance equation are also included. In the SCUC model, the line thermal limit for fixed line rating during a day is shown in (29).

$$-P_k^{max} \leq P_{kt} \leq P_k^{max} \quad \forall k,t \tag{29}$$

The nodal power balance constraint (30) included in the SCUC considers renewable curtailment.

$$\sum_{g\in G(n)} P_{gt} + \sum_{k\in K(n-)} P_{kt} - \sum_{k\in K(n+)} P_{kt} + \sum_{w\in R(n)} P_{rt} - \sum_{w\in R(n)} P_{rt}^{Cur} = d_{nt} \quad \forall n,t \tag{30}$$

The curtailment variable $P_{rt}^{Cur}$ is constrained by (31).

$$0 \leq P_{rt}^{Cur} \leq P_{rt} \quad \forall r,t \tag{31}$$

Here $P_{rt}^{Cur}$ lets the optimiser curtail renewable unit $r$ at time $t$ as needed, while $P_{rt}$ is the uncurtailed availability we derived in our TX-123BT dataset.

The SCUC simulations are conducted to verify daily system profiles of the TX-123BT test case with daily line ratings. Since the renewable production, line rating and loads vary differently in the 5-year period, the feasibility of SCUC optimization problems can imply the created test system is practical and reliable.

### *B. Comparison of the Electricity Market*

Since the locational marginal prices (LMPs) can be obtained from the SCUC simulations, the electricity market results of the TX-123BT and actual ERCOT can be compared.

#### *1) Actual ERCOT Electricity Market*

Electricity prices are affected by many factors and one year's price data may not well reflect the actual electricity market. Hence, we collect and analyze the day ahead market (DAM) price data in a 5-year period. After observing DAM prices for different hours and load zones under different scenarios, some characteristics of the actual ERCOT electricity prices are observed and summarized as follows,

- Electricity prices on weekends are usually lower than the prices on weekdays.
- Quarter 3 has the highest electricity price while Quarter 1 has the lowest electricity price.
- The electricity prices at different load zones are slightly different during off-peak hours, but the electricity prices usually have larger locational variety during peak hours, especially in Quarter 3.
- For the peak hours around 15:00-18:00, the electricity prices are much higher than the off-peak hour prices in Quarter 3.

#### *2) Synthetic TX-123BT Electricity Market*

The LMPs of the TX-123BT are obtained using the dual variables of the nodal power balance constraints in SCUC simulation. After the analysis of the TX-123BT LMPs and the ERCOT DAM prices, we conclude that the two systems have very similar nodal electricity prices range under different scenarios, which is shown in Table VI.

TABLE VI Electricity Price Range ($/MWh) Under Different Scenarios

| Scenario | | Trough Hour | Normal Hour | Peak Hour |
|---|---|---|---|---|
| Quarter 1 | Weekday | 15-24 | 24-30 | 30-45 |
| | Weekend | 15-20 | 20-25 | 25-33 |
| Quarter 2 | Weekday | 16-25 | 25-35 | 35-55 |
| | Weekend | 16-25 | 25-30 | 30-50 |
| Quarter 3 | Weekday | 18-25 | 25-50 | 50-220 |
| | Weekend | 18-25 | 25-45 | 45-110 |
| Quarter 4 | Weekday | 17-25 | 25-30 | 30-40 |
| | Weekend | 15-24 | 24-28 | 28-45 |

The system-wide electricity prices for different typical seasonal days are shown in Figs. 10a-10d. We can observe that, in Quarters 3, the electricity prices are higher than in Quarters 1. This disparity can be explained by the larger demands in Quarters 3. The high demands require generators which are more expensive for electricity production to come online, resulting in higher electricity prices in these quarters.

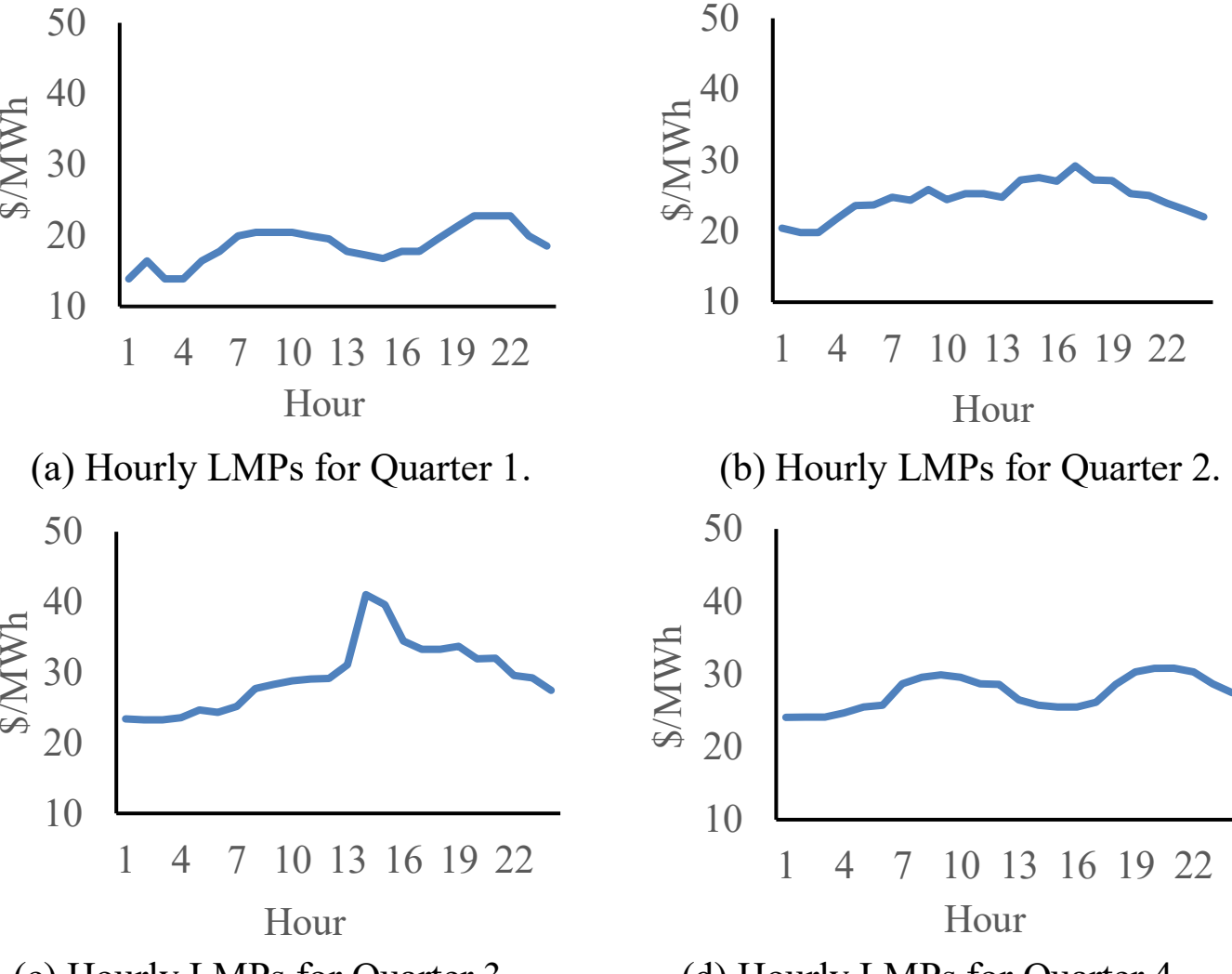

(a) Hourly LMPs for Quarter 1. (b) Hourly LMPs for Quarter 2.

(c) Hourly LMPs for Quarter 3. (d) Hourly LMPs for Quarter 4.

Fig. 10. Daily electricity prices for different quarters.

Previously, Fig. 9 shows that the line 15 dynamic rating dips significantly at midday (due to high temperature and solar heating of the conductor). This means the grid's transmission capacity is lowest precisely when solar generation is at its peak. The SCUC simulation reflects this constraint – we observed that in Q3, the highest hourly LMP occurred during the hour of lowest line rating. In other words, at the peak solar hour on a hot summer day, line's capacity was most limited, leading to congestion and a sharp LMP increase.

The day with the highest load among all the days is selected as the peak load day. Two scatter plots of nodal LMPs for the normal load day and peak load day are shown in Fig. 11-12. From the simulation results, we can conclude that the electricity prices at different load zones are slightly different during low load demand scenarios (for most buses). However, the electricity prices locational variety is large during peak hours in Quarter 3. The characteristics of the LMPs are in line with the actual ERCOT electricity price characteristics that we summarized in the above subsection.

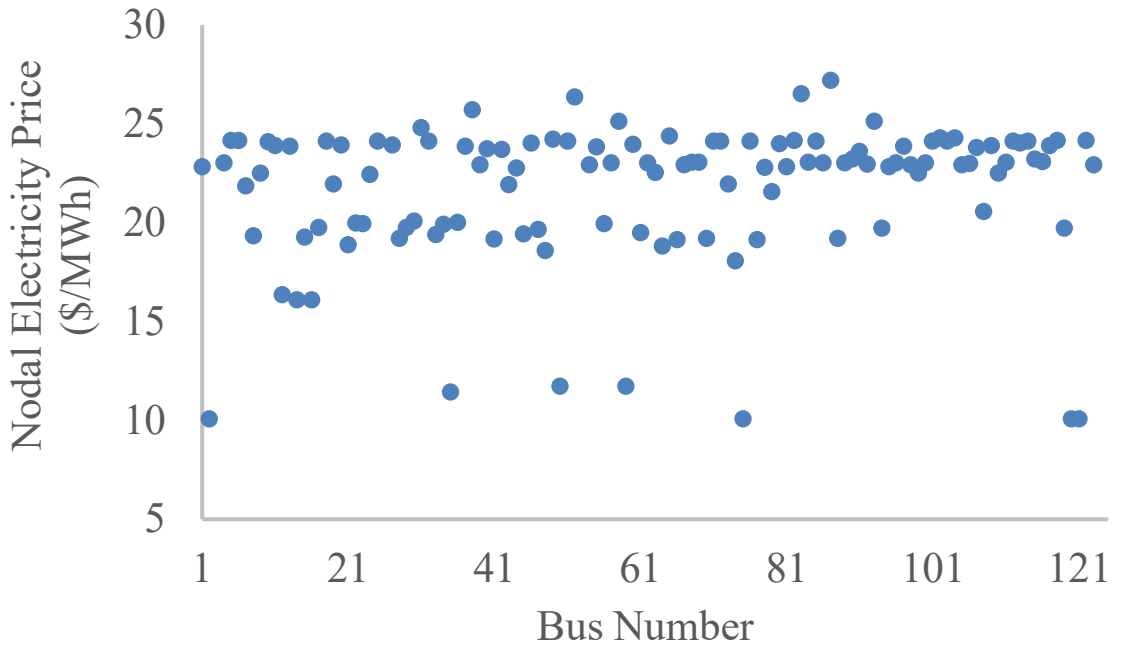

Fig. 11. Nodal LMPs for Hour 15 in a normal load day.

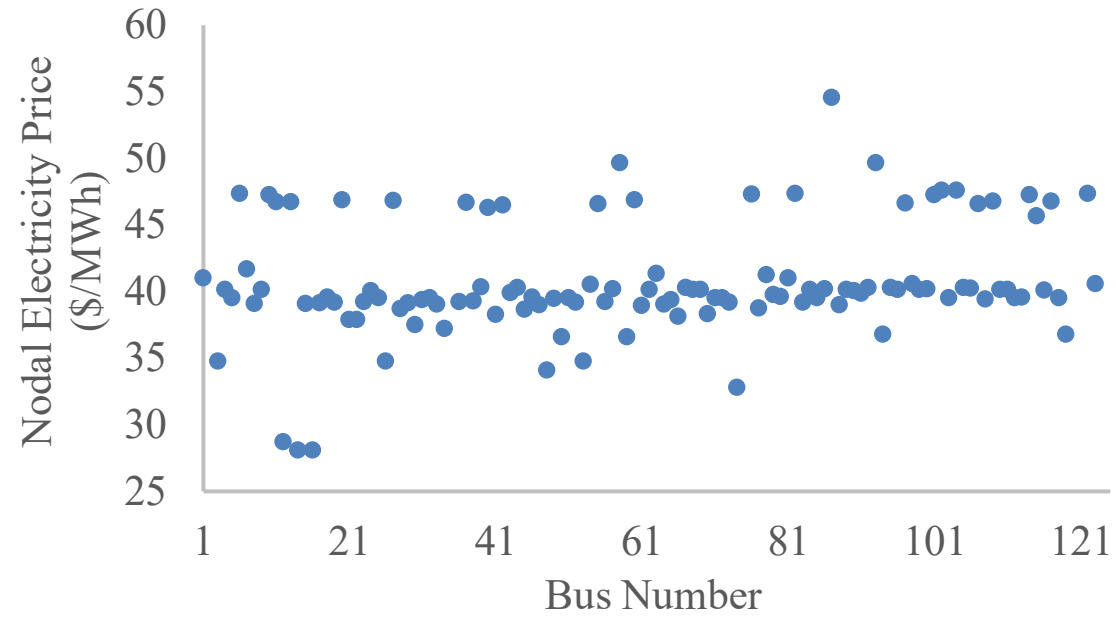

Fig. 12. Nodal LMPs for Hour 15 in the peak load day.

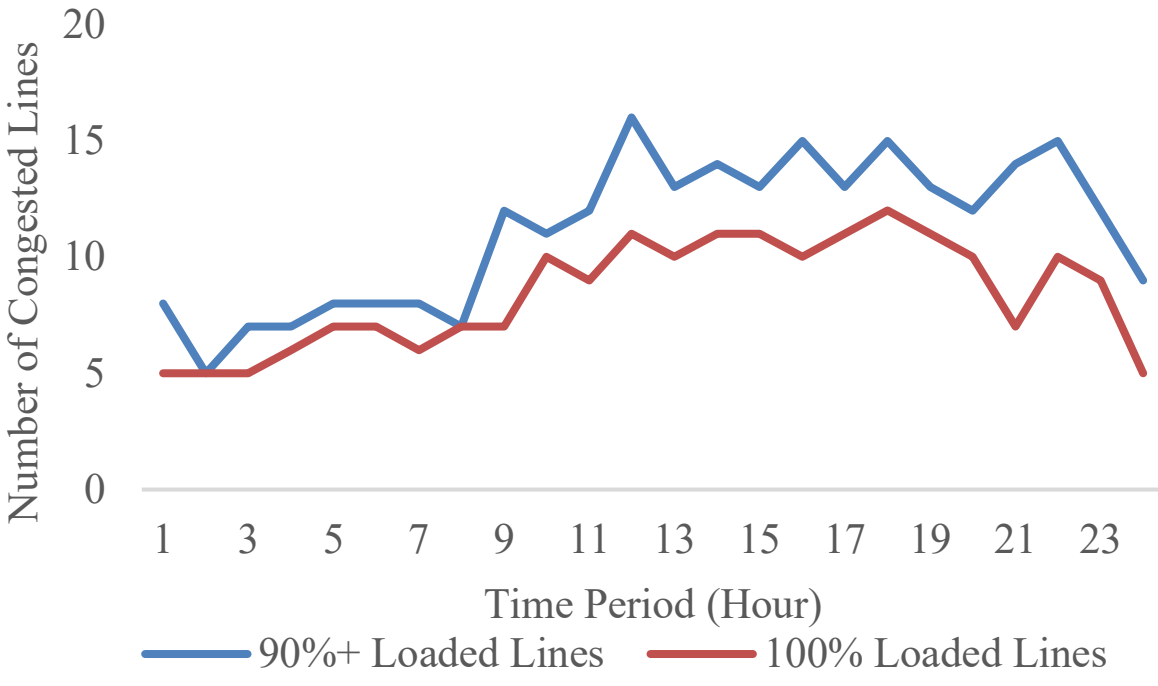

Fig. 13. Number of congested lines during the peak load day.

## C. Peak Load Scenarios and Line Congestion Analysis

Congested lines are classified based on the power flow results of the SCUC simulations. Two types of congested lines are classified: (i) 100% loaded lines and (ii) 90%+ loaded lines. The 100% loaded lines are the transmission lines on which the active power flow is 100% of the line capacity. The 90%+ loaded lines are the transmission lines on which the active power flow is over 90% but less than 100% of the line capacity. The numbers of congested lines at different hours during the peak load day are shown in Fig. 13. We can observe more transmission lines are congested in the peak hours.

## D. DLR Performance Analysis

The SCUC simulation is also conducted on the TX-123BT with hourly DLR profiles. The line thermal limit constraint is replaced by the following constraint since the line limits $P_{kt}^{max}$ are now different for different hours and need an extra index of time interval $t$.

$$-P_{kt}^{max} \leq P_{kt} \leq P_{kt}^{max} \quad \forall k, t \tag{32}$$

The SCUC simulation results including total operational cost, renewable generation, LMPs, and transmission congestion, are analyzed and compared with SCUC using daily DLR profiles. The overall numerical results are shown in Table VII.

TABLE VII SCUC Simulation Results for a Normal Day in Quarter 2

| Numerical Results | Daily DLR | Hourly DLR |
|---|---|---|
| Total Operational Cost ($) | 8.09M | 7.95M (-1.7%) |
| Total Renewable Generation (GWh) | 271.95 | 275.48 (+1.3%) |
| Average LMPs ($/MWh) | 18.66 | 17.98 (-3.6%) |
| ANCLPH | 6.9 | 8.2 |

*ANCLPH denotes the average number of congestion lines per hour.

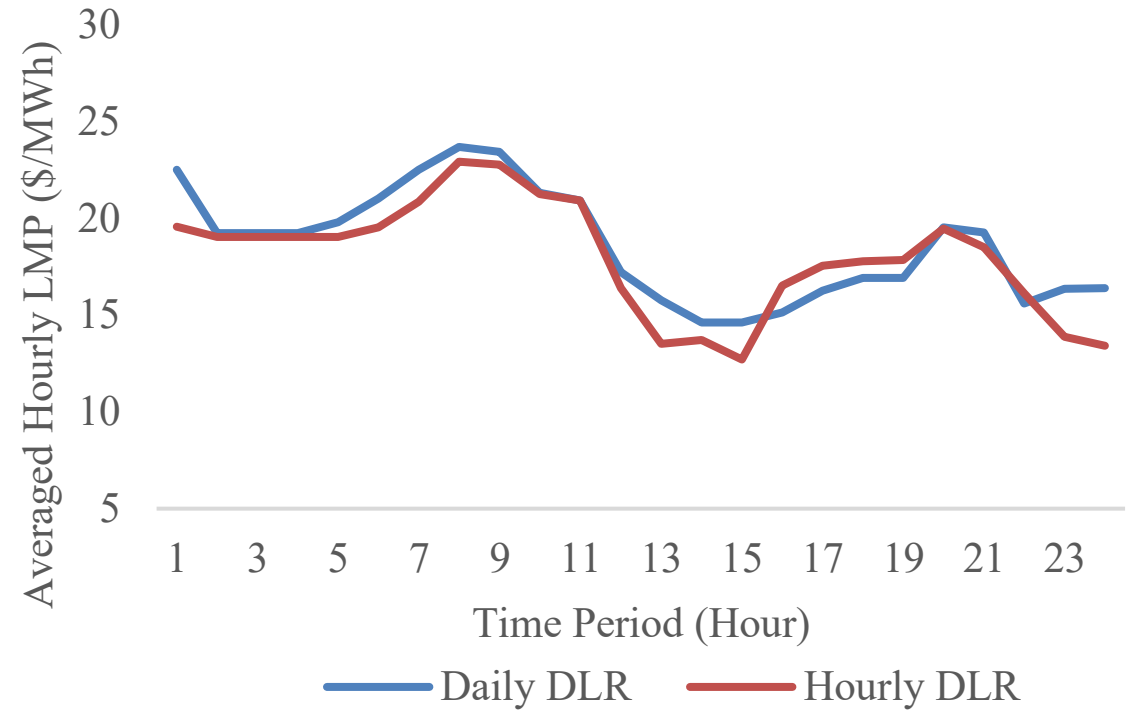

Fig. 14. Hourly average LMPs for a normal day in Quarter 2.

The total operational cost of the hourly DLR case is lower than the daily line rating case, and the cost saving is about 1.7% with hourly DLR. One reason is that the increased transmission capacity can relieve network congestion and reduce the curtailment of renewable energy that has a much lower (zero) cost than the conventional generation. The average LMP of the hourly DLR case is also lower than the case using conservative daily DLR. The systemwide average LMPs for a normal day in Quarter 2 are shown in Fig. 14. We can observe that the average LMPs of hourly DLR is lower than the LMPs of daily DLR for majority of the hours.

# VIII. Hydrogen Studies on TX-123BT

The TX-123BT test case features spatio-temporal renewable production profiles. These profiles are particularly valuable for studies requiring detailed, long-term data on renewable energy production which address its versatility. In this section, we explored potential hydrogen facility investments in TX-123BT. We examined two possible integration configurations: a hydrogen hub (HB) and a Hydrogen Energy Transmission and Conversion System (HETCS). We analyze how the TX-123BT operates with these hydrogen facilities and further investigate how the level of detail in renewable production profiles affects the operation performance.

## A. Hydrogen Investments in TX-123BT

The hydrogen hub is a localized plant which includes hydrogen tank, electrolyzers and fuel cells [47]. This setup allows for electrical energy to be converted and stored locally, then converted back when needed. The hydrogen hub can realize energy storage functions from the grid's perspective and be deployed near renewable resources to mitigate the intermittency and uncertainty associated with renewable production. Within the territory of TX-123BT, we have deployed two hydrogen hubs close to areas of renewable production. The incorporation of these two potential hydrogen hubs into TX-123BT is henceforth referred to as TX-123BT HB.

The Hydrogen Energy Transmission and Conversion System (HETCS) encompasses hydrogen pipelines, with electrolyzers and fuel cells installed at the terminal ends of the pipelines [48]. This configuration enables the conversion and transmission of electrical energy in the form of hydrogen for an HETCS-integrated grid. HETCS can be particularly advantageous in grids with high renewable energy penetration for the long-distance transmission of surplus renewable energy. Consequently, we have deployed two HETCS lines extending from the renewable resource areas located in the northeast of Texas to the cities in the southwest. The integration of HETCS into TX-123BT is henceforth referred to as TX-123BT HT.

### *B. Operation of TX-123BT with Hydrogen Facilities*

For grids integrated with HB or HETCS, grid operation should consider not only the scheduling and dispatch of generators but also the working status of related hydrogen facilities, such as fuel cells, electrolyzers, hydrogen pipelines, and storage units. Co-optimizing the operations of both electrical and hydrogen facilities can maximize the overall benefits from a grid perspective. The enhanced SCUC models for hydrogen-integrated grids [47] are utilized for daily operation simulations on TX-123BT HB and HT scenarios. These grid operation conditions are compared with those of TX-123BT without any hydrogen facilities, with findings presented in Table VIII. Across all four quarters, the inclusion of hydrogen facilities consistently reduces daily operation costs. This confirms the specific locations and configurations of HB and HT are economically advantageous. Depending on the daily grid conditions, either the HB or HT configuration may offer the lowest daily operation costs.

TABLE VIII Daily Operation Costs of TX-123BT with Various Hydrogen Integration in Different Quarters (M$)

| Quarter | No Hydrogen Case | HB Case | HT Case |
|---|---|---|---|
| Quarter I | 16.748 | 16.728 | 16.722 |
| Quarter II | 7.528 | 7.508 | 7.537 |
| Quarter III | 22.714 | 19.546 | 19.506 |
| Quarter IV | 11.267 | 11.240 | 11.229 |

### *C. Evaluation on the Temporal Resolution of Profiles*

The TX-123BT dataset is distinguished by providing hourly profiles for renewable production and line ratings, which other datasets do not offer. To investigate the necessity of these high temporal resolution profiles and their effect on power system operations, SCUC simulations were performed on the HB and HETCS cases under the assumption that these spatio-temporal profiles were not accessible. The results were then contrasted with the SCUC results that included these profiles.

TABLE IX Comparison of Daily Operation Costs (M$) for Hydrogen Studies Utilizing TX-123BT DLR Profiles

| Quarter | HB with HLRP | HB without HLRP | HT with HLRP | HT without HLRP |
|---|---|---|---|---|
| Quarter I | 16.728 | 17.483 | 16.722 | 17.536 |
| Quarter II | 7.508 | 8.838 | 7.537 | 8.028 |
| Quarter III | 19.546 | 22.502 | 19.506 | 21.824 |
| Quarter IV | 11.240 | 11.723 | 11.229 | 11.729 |

*HLRP represents hourly line rating profiles

In Table IX, the costs of daily operation with and without the dynamic line rating information are compared. It is inferred that the absence of hourly line rating data compromises the ability to accurately gauge the economic advantages of integrating hydrogen. Specifically, without access to DLR profiles, the costs associated with daily operations can surpass those incurred prior to hydrogen integration. This suggests the critical role of DLR information in optimizing economic benefits in energy systems integrating hydrogen.

TABLE X Comparison of Daily Operation Costs (M$) for Hydrogen Studies Utilizing TX-123BT Weather-Dependent Renewable Production Profiles

| Quarter | HB Case with WRP | HB Case without WRP | HT Case with WRP | HT Case without WRP |
|---|---|---|---|---|
| Quarter I | 16.728 | 20.888 | 16.722 | 20.887 |
| Quarter II | 7.508 | 12.832 | 7.537 | 12.830 |
| Quarter III | 19.546 | 19.546 | 19.506 | 19.506 |
| Quarter IV | 11.240 | 15.399 | 11.229 | 15.466 |

*WRP represents weather-dependent renewable production profiles

The daily operation costs, both with and without considering the weather-dependent renewable production profiles, are presented in Table X. Both the HB and HT cases, without the inclusion of hourly renewable production profiles based on the weather information, result in higher operation costs. However, an exception is noted in Quarter III, where loads are high and most renewable energy sources are utilized. In such situations, the detailed information on renewable production becomes less significant.

## IX. Conclusion

This paper details the creation of Texas 123-bus backbone transmission system, a synthetic test system that spans the Texas grid footprint. The created test case has high-voltage backbone transmission network while retaining geographical characteristics. Hence, the test case is suitable for power system studies that require geographical information and less computational burden. The hourly weather data Phase 2 of North American Land Data Assimilation System, including solar radiation, air temperature and wind speed for all the 123 bus locations are extracted and utilized. Using the weather-dependent models for solar/wind production and transmission line rating, the associated spatio-temporally correlated profiles are created. The nodal load profiles are also created in a way to match the actual zonal load for each of the eight weather zones in Electric Reliability Council of Texas. Since the created profiles covers a 5-year period at one-hour resolution with strong practical spatiotemporal correlations embedded in the dataset, it would also facilitate power system studies involving machine learning such as fully-connected neural networks and graph neural networks [49]-[50].

The security-constrained unit commitment simulation results demonstrate that the Texas 123-bus backbone transmission system can closely replicate the actual Electric Reliability Council of Texas market behavior over multiple years, validating the realism of the synthetic dataset. Moreover, the observed 1.7% reduction in operational costs when using hourly dynamic line ratings (as opposed to daily ratings) illustrates a tangible benefit of incorporating high-resolution temporal data — an insight that can inform real-world grid operation and planning. Additionally, the hydrogen integration case study underscores the value of spatio-temporal profiles in evaluating future energy solutions, as it reveals how spatio-temporal renewable data can influence infrastructure investment decisions. The security-constrained unit commitment, network congestion and hydrogen integration studies demonstrate the effectiveness and practicality of the created synthetic system for facilitating research and studies in various power system areas.